# PULSE: The Palomar Ultraviolet Laser for the Study of Exoplanets

Christoph Baranec*[a], Richard G. Dekany[b], Rick S. Burruss[c], Brendan P. Bowler[d], Marcos van Dam[e], Reed Riddle[b], J. Christopher Shelton[c], Tuan Truong[c], Jennifer Roberts[c], Jennifer Milburn[b] & Jonathan Tesch[c]

[a]Institute for Astronomy, University of Hawai'i at Mānoa, Hilo, Hawai'i 96720-2700 USA; [b]Caltech Optical Observatories, California Institute of Technology, Pasadena, California, 91125 USA; [c]Jet Propulsion Laboratory, California Institute of Technology, Pasadena, California 91109, USA; [d]Division of Geological and Planetary Sciences, California Institute of Technology, Pasadena, California, 91125 USA; [e]Flat Wavefronts, Christchurch, 8140, New Zealand

## ABSTRACT

The Palomar Ultraviolet Laser for the Study of Exoplanets (PULSE) will dramatically expand the science reach of PALM-3000, the facility high-contrast extreme adaptive optics system on the 5-meter Hale Telescope. By using an ultraviolet laser to measure the dominant high spatial and temporal order turbulence near the telescope aperture, one can increase the limiting natural guide star magnitude for exquisite correction from $m_V < 10$ to $m_V < 16$. Providing the highest near-infrared Strehl ratios from any large telescope laser adaptive optics system, PULSE uniquely enables spectroscopy of low-mass and more distant young exoplanet systems, essential to formulating a complete picture of exoplanet populations.

## 1. INTRODUCTION

Direct imaging of exoplanets is a burgeoning field of astrophysics recently enabled by a second-generation of adaptive optics (AO) systems optimized for high-contrast observations around bright stellar hosts[1-4]. PALM-3000, the facility high-contrast extreme adaptive optics system on the 5-meter Hale Telescope, which started operations in June 2011, was the first multi-stage adaptive optics system combining exoplanet imaging techniques such as coronagraphy[5], non-common-path electric field minimization[6], and integral field chromatic differential speckle imaging[7,8]. It is now actively providing direct infrared spectra of nearby giant exoplanets[8], and the sharpest visible-light images from a single aperture[1]. Despite this success, current adaptive optics guide star brightness requirements preclude study of low-mass M type stellar dwarfs as well as the youngest stars in star forming associations out to 200 pc. It is essential to forming a complete picture of exoplanet development to bring these populations of stars and exoplanets under study.

The Palomar Ultraviolet Laser for the Study of Exoplanets (PULSE) will dramatically expand the science reach of PALM-3000. By reusing the reliable ultraviolet Rayleigh laser guide star successfully deployed by Robo-AO on the Palomar 1.5-m telescope[10], we will open the sky to near-infrared high-contrast and diffraction-limited visible light observations by increasing the limiting natural guide star magnitude for exquisite correction from $m_V < 10$ to $m_V < 16$. Simulations indicate that very-high infrared contrast ratios and good visible-light adaptive optics performance will be achieved by such an upgraded system on stars as faint as $m_V < 16$ using an optimized low-order NGS sensor. PULSE uniquely enables spectroscopy of low-mass and more distant young exoplanet systems, data essential to formulating a complete picture of exoplanet populations. Beyond exoplanet work, PULSE will have wide science impact by increasing the available fraction of sky accessible with Palomar AO from ~1% to 50%. Enabled science includes visible-light diffraction limited studies of Kuiper Belt Object multiplicity and composition, near-infrared gas and dust emissions in stellar environments, tracing dark matter in globular cores, astrometric follow-up of Gaia targets, velocity dispersion of galaxies hosting active nuclei, and spatially-resolved velocimetry of merging galaxies. With the highest near-infrared Strehl ratios from any large telescope laser adaptive optics system and a suite of 5 back-end instruments, PULSE will enable an exceptional range of science.

*baranec@hawaii.edu; phone 1 808 932-2318; http://high-res.org

## 2. THE NEED FOR PULSE

Beginning 2014, exoplanet-targeting AO systems, such as the Gemini Planet Imager (GPI) and Spectro-Polarimetric High-contrast Exoplanet REsearch (SPHERE), will be commissioned on 8-10 meter class telescopes[2-4,11]. These systems are expected to push toward direct imaging contrast of $10^{-7}$, but will be similarly restricted to use NGS of $m_V < 10$ for high-contrast performance. No exoplanet-optimized AO system plans laser support and none has the instrument suite of PULSE (see Table 1). Unlike any other system, PULSE will deliver Strehl ratio > 50% in H-band (1.65 μm wavelength) for NGS of $m_V \sim 15$ and in K-band (2.2 μm) for $m_V \sim 16$ in median conditions (Figure 1). 50% H-band Strehl is crucial as experience has shown this performance is necessary for effective stellar rejection through coronagraphy. PULSE also opens access to diffraction-limited science (> 10% Strehl) in red optical bands for $m_V \sim 14.5$. In 25th percentile Palomar seeing (0.8" full-width at half-maximum; FWHM), both H-band coronagraphy and 750 nm-band science are enabled for ~ 1 magnitude fainter guide star.

Existing sodium laser guide star (LGS) AO systems on 8-10 meter telescopes, now workhorses for astronomy[12,13], were conversely designed for low-Strehl general-purpose NIR science; a similar, earlier technology demonstration of sum-frequency sodium LGS technology at Palomar has been successfully concluded[14]. LBT exoplanet science is targeting NGS AO in the thermal infrared, while its future LGS system is optimized for wide-field moderate correction[15] ground-layer AO. MMT and Magellan are not outfitted for NIR direct exoplanet spectroscopy.

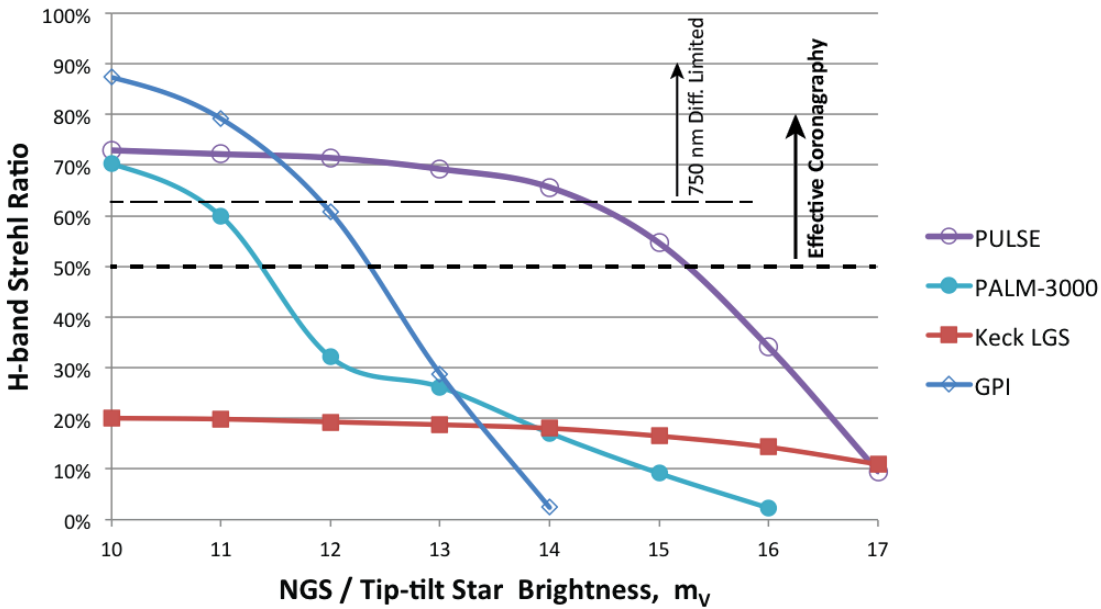

Figure 1. H-band Strehl ratio for various AO systems. PULSE reaches guide stars much fainter than GPI, at correction levels far superior to Keck LGS AO - which never exceeds 20% H-band Strehl. The PALM-3000 curve reflects its three selectable wavefront sensor modes. All curves are for median seeing at their respective sites.

The UV LGS AO SAMS system on the 4-meter SOAR telescope[16] is similarly optimized for wide-field AO and not direct exoplanet spectroscopy. The 3-meter ShaneAO system[17] will be another general-use IR system and sodium LGS demonstrator, lacking the site quality, aperture, and high-contrast infrastructure of PULSE. Finally, Robo-AO itself is limited in aperture for direct exoplanet imaging.

The Hubble Space Telescope (HST) is the world's superior resource for diffraction-limited visible imaging[18], but, for modest fields around guide stars, PULSE will surpass by twice HST's 2.4-meter imaging resolution, and do so with readily upgradable instruments. Proposed small-aperture exoplanet imaging space missions[19-21] will not have the collecting area nor versatility of PULSE for resolved spectroscopy.

Eventually, extremely large telescope (25-30 meter diameter) projects will enter PULSE's discovery space. Their general-purpose LGS AO will compete in imaging contrast, but these observatories are not expected until after 2020, with speckle-suppressing exoplanet instruments[22] expected closer to 2030. Lessons from LGS + NGS validation are likely to inform designs for multi-sodium laser + NGS hybrids.

Table 1. Current and future instruments available to PULSE, including NIR and visible imagers, slit spectrographs, and integral field spectrographs (IFS).

| | PULSE Instrument | Type | Wavebands | Max R | FoV | New PULSE Science Programs Enabled by Increased Sky Coverage |
|---|---|---|---|---|---|---|
| **Operational** | **PHARO** | NIR Imager, Spectrograph, Coronagraph | Y, J, H, K | ~ 1500 | 40" × 40" | Small IWA coronagraphy in young associations; Precision astrometry of globular cluster cores; IMBHs; Circumstellar dust and gas spectroscopy; ULIRGs, SMGs, QSOs, strong gravitational lenses |
| | **Fiber Nuller** | Two-Aperture NIR Nulling Interferometer | J, H, K | ~ 5 | N/A | Very close orbit (< 2 λ/D separation) NIR exoplanet search at ~$10^{-3}$ contrast |
| | **SWIFT** | Visible IFS | 630 – 1050 nm | ~ 4000 | 10" × 20" | Resolved synoptic planetary IFS spectroscopy; SNe kinematics; Cluster IMFs; H-alpha emissions; Galaxy mergers, disk formation, dark matter kinematics; AGN hosts to z ~ 1.5 |
| | **Project 1640** | NIR Coronagraph + IFS | J – H simultaneous | ~ 35 | 4" × 4" | Nearby M dwarf direct exoplanet spectroscopy, exoplanet atmospheres, composition, dust at $10^{-6}$ contrast |
| | **TMAS** | Visible High-Speed Imager | B – I + misc. narrow filters | ~ 300 (narrow filters) | 25" × 20" | Comets, asteroids, Galilean satellites, KBOs; White dwarf companions; Supergiant starspots; Star-formation in Orion; Herbig-Halo objects and jets |
| **Funded** | **Stellar Double Coronagraph** | NIR/Visible Coronagraph | B – K | Instrument-based | 60" × 60" | Self-calibrating two-stage vector vortex coronagraph for high-contrast use with each of PHARO, TMAS, and DARKNESS |
| | **DARKNESS** | Energy-resolving, photon-counting MKIDS Camera | 700 – 1350 nm | ~ 30 | 2.5" × 2.5" | High-contrast speckle suppression via photon arrival statistics; Time-domain astrophysics via photon time-tagging; Debris disks; Galactic microlensing events |

PULSE, therefore, is uniquely positioned to open the realm of precision wavefront, faint-guide-star science from the ground. It leverages a proven high-contrast capability built of ~ $12M of instrumentation investment, 3 years of direct Robo-AO experience with LGS operations, and an accessible, moderate altitude (5500') site conducive to integrating research and student learning.

PULSE will be available to > 400 astronomers at Caltech, Yale University, American Museum of Natural History (New York), Jet Propulsion Laboratory, Oxford University, and NAOC.

## 3. PULSE SYSTEM DESCRIPTION

### 3.1 Architecture

PULSE will reuse the proven Robo-AO ultraviolet (UV) Rayleigh laser guide star to form a brilliant (equivalent to $m_V$ ~ 7), reliable, pilot-safe source for rapid wavefront sensing (Figure 2). Using the same PULSE laser hardware, Robo-AO has performed over 12,000 science observations at the visible-light diffraction limit of the 1.5-m telescope with exemplary robustness and efficiency[23]. Though we draw from our Rayleigh LGS experience, the technology is mature and has been deployed successfully for modest levels of AO correction at SOAR, WHT, MMT, and Mt. Wilson Observatories.

The primary challenge to using a low-altitude Rayleigh LGS on a 5-meter diameter aperture arises from the geometric configuration of the beacon, which is range-gated to sample light returned from 10 km, where Rayleigh backscatter remains strong. This modest beacon altitude results in a geometric discrepancy between the (conical) wavefront sensing path and (cylindrical) atmospheric propagation path for collected light from astronomical sources. The magnitude of root-mean-squared (RMS) wavefront sensing error arising from this focus anisoplanatism (FA) effect is given by,

$$\sigma_{LGS}^{FA} = \left(\frac{D}{d_{0[0.5\,\mu]}}\right)^{5/6} \left(\frac{0.5\,\mu}{\lambda}\right) \text{ [nm]} \qquad (1)$$

where D is the telescope diameter and d0 is the focus anisoplanatism coherence parameter[24]. For D = 5m and median atmospheric conditions measured at Palomar[25], our (optimum) 10 km altitude beacon would result in $\sigma_{LGS}^{FA} \sim 360$ nm wavefront error, far too large for high-contrast observations (capping performance from this one error term at 15% Strehl ratio in H-band and essentially none in the red optical). For efficient coronagraphy, we seek from experience a total wavefront error from all AO error sources less than 220 nm RMS (corresponding to 50% Strehl in H-band.) For 750nm-band science, wavefront error less than 180 nm RMS is appropriate (equivalent to 62% H-Strehl).

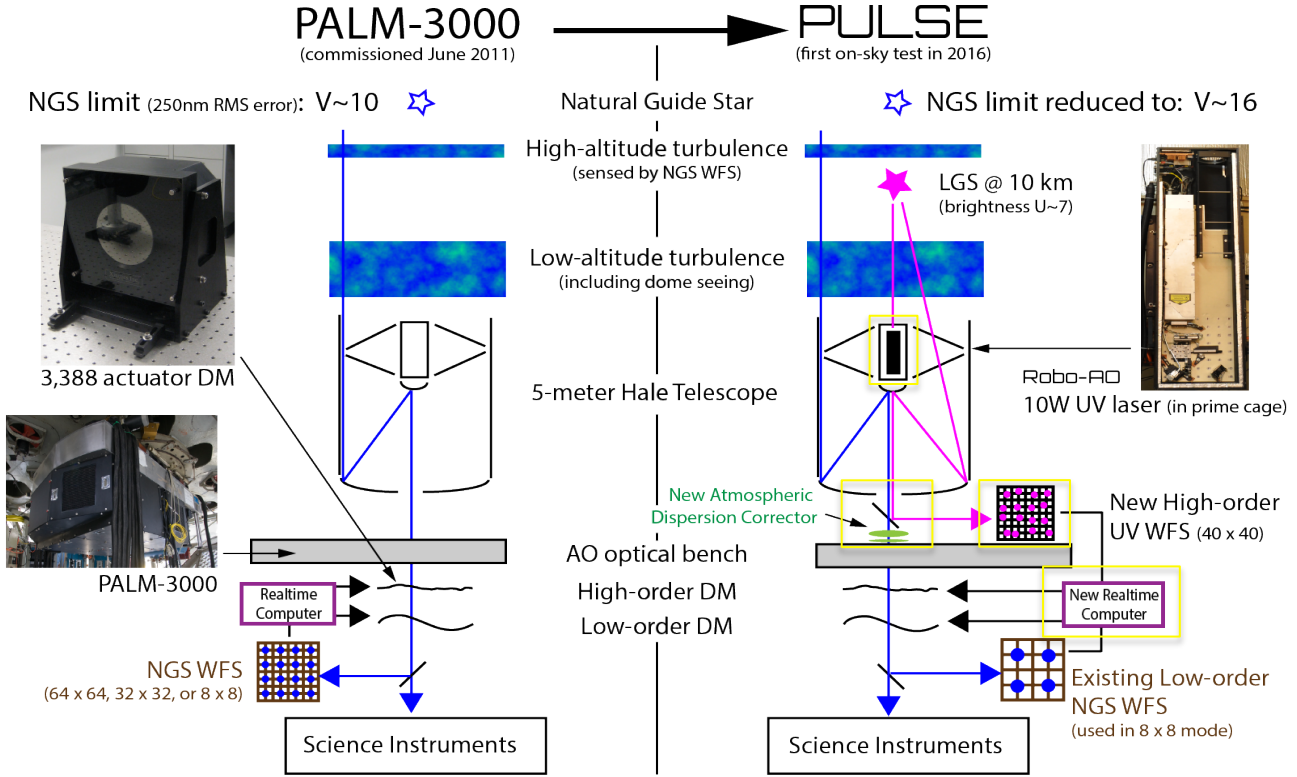

Figure 2. Overview of proposed upgrade path and system architecture for PULSE (discussed in the text.) The laser projector shown at right is the actual hardware to be re-used for PULSE. A second, somewhat less commonly considered limitation of Rayleigh LGS wavefront sensing arises from the fact that some atmospheric turbulence is present in stratospheric layers above 10km altitude. Fortunately, detailed consideration of high-altitude turbulence at Palomar49 has shown that the characteristic turbulence scale, given by the Fried parameter r0, corresponding to turbulence above 10 km is approximately 0.6 meter (for 500 nm wavelength light). PULSE's low-order NGS sensor, having a subaperture diameter of 0.64-meters (§3.3), is thus very well equipment to accurately measure that turbulence arising above the Rayleigh LGS.

To overcome the focus anisoplanatism, we first recognize that the manifestation of focus anisoplanatism error is dominated by low-spatial-frequency residual aberrations, and we augment our Rayleigh LGS sensing with a modest number of spatial wavefront modes measured directly from a natural guide star. Although the need for NGS tip-tilt mode sensing with any LGS AO is well established[26], the hybridization of LGS and NGS sensing of multiple low-order modes is a key innovation, borne out by detailed AO simulations (§3.2). Intuitively, the efficacy of this solution can be understood by considering the FA error to arise primarily from the perimeter of the telescope pupil where the geometric LGS error is large; interior to the pupil little geometric discrepancy due to finite beacon altitude exists. Because a 5-meter telescope requires augmentation of the UV LGS with relatively few spatial modes (~ 50), the corresponding NGS faintness limit can be significantly extended, over 100x fainter, for comparable performance.

A second, somewhat less commonly considered limitation of Rayleigh LGS wavefront sensing arises from the fact that some atmospheric turbulence is present in stratospheric layers above 10km altitude. Fortunately, detailed consideration of high-altitude turbulence at Palomar[25] has shown that the characteristic turbulence scale, given by the Fried parameter $r_0$, corresponding to turbulence above 10 km is approximately 0.6 meter (for 500 nm wavelength light). PULSE's low-order NGS sensor, having a subaperture diameter of 0.64-meters (§3.3), is thus very well equipped to accurately measure that turbulence arising above the Rayleigh LGS.

## 3.2 Adaptive Optics Performance Simulation Results

We have performed detailed Monte Carlo wave-optics propagation and control simulations[27] of PULSE using yao, a mature AO performance simulation package widely used in the AO community[28]. From these, we have developed a parametric performance model that includes all significant error terms: focus anisoplanatism, finite beacon height, measurement noise, deformable mirror fitting error, servo control bandwidth, scintillation effects, and non-common path calibration aberrations.

For our selected architecture consisting of 40 × 40 LGS wavefront sensing samples operating at 1 kHz frame rate, and our existing 8 × 8 subaperture NGS wavefront sensor[1], we find performance predictions (for a simulated 10-second science integration) summarized in Table 2. Therein, we see that excellent high-contrast correction can be obtained in median 1.1" FWHM seeing conditions to $m_V \sim 15$ and usable K-band correction to $m_V \sim 17$. For 25% percentile (best) seeing of 0.83" FWHM, similar analyses show the 80% K-Strehl limit at $m_V \sim 16.1$ and 15% K-Strehl limit at an impressive $m_V \sim 18.3$. At these faintness limits, exploiting the PALM-3000 patrol range for acquiring an NGS (an ellipse approximately 90 × 120 arcseconds), some 50% of the available Northern sky becomes accessible to high-quality AO correction with PULSE, with near complete coverage at low galactic latitudes[26].

Table 2. Monte Carlo-based performance prediction as a function of NGS brightness. In all cases, the LGS loop samples 40 × 40 wavefront subapertures at 1kHz, using photoreturns confirmed on-sky with Robo-AO, having corresponding subaperture size.

| Median K-band Strehl Ratio | NGS Wavefront Sensor Sample Rate | | |
|---|---|---|---|
| NGS brightness, $m_V$ | 1 kHz | 250 Hz | 100 Hz |
| 10 | **84%** | -- | -- |
| 14 | **78%** | 75% | 52% |
| 15 | -- | **67%** | 49% |
| 16 | -- | **51%** | 40% |
| 17 | -- | 10% | **22%** |

The comparison of performance between PULSE and PALM-3000 in several observing modes and median conditions is shown in Figure 3. In all cases we consider Strehl ratio in K-band for uniform comparison. We see that PALM-3000 performance rolls off steeply at $m_V \sim 9$ when sampling with 64 subapertures across the pupil diameter (8.2 cm each), as insufficient photons are available within an atmospheric coherence time for accurate wavefront sensing. (Not shown, performance using 32 subapertures across the pupil (16.4-cm each) rolls over at $m_V \sim 10$). Our NGS mode with 8 subapertures across (63.8-cm each) reaches reasonable faintness, but performance is limited by wavefront fitting error to low Strehl ratios, no better than 25% at $m_V \sim 15$. If one were to interpret the advantage of PULSE as only removing the fitting error term of our 63.8-cm subaperture mode, performance would follow the dashed curve in Figure 5.

However, a key benefit arises with PULSE: the NGS sensor subimages are sharpened by the action of the LGS control loop, boosting performance to the PULSE curve shown. This is comparable to the NGS faintness advantage of tip-tilt star sharpening well-known with sodium LGS systems[29], except that here we do not expect diffraction-limited subimages. Rather models suggest we may see ~ 0.4" FWHM images within the Shack-Hartmann sensor, a modest but important gain over 1.1" seeing, allowing over 60% Strehl ratio for $m_V = 15$.

We further minimize cost and risk by noting that the 8 × 8 subaperture NGS sensing system required for PULSE, in fact, already exists in PALM-3000 as one of three pupil sampling modes (64, 32, or 8 subapertures per pupil diameter). This fortunate situation significantly reduces development cost and project risk and is our baseline.

Our simulations suggest Rayleigh LGS are not ideal for telescopes larger than 5-meter diameter. As telescope size increases, wavefront error variance from focus anisoplanatism grows as $D^{5/3}$ greatly increasing required measurement dynamic range and linearity requirements. Our approach, however, may indeed be very relevant to higher-beacon-altitude sodium LGS AO systems on 8-10 meter telescopes currently limited by focus anisoplanatism, such as Keck LGS

AO. Experience with PULSE will verify for the AO community both performance and feasibility of our hybrid sensing approach. However, because Rayleigh LGS provide over 10x the wavefront sensing photo-return of sodium LGS per laser Watt – and over 100x the photo-return per dollar – PULSE will remain competitive for the highest AO performance applications for the foreseeable future.

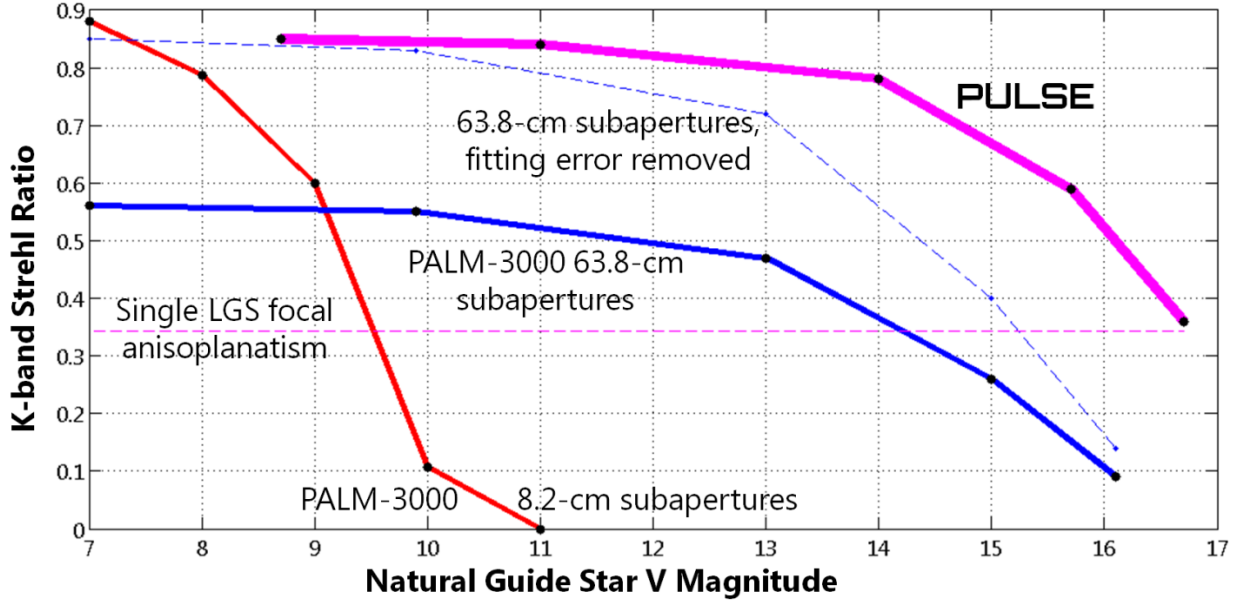

Figure 3. K-band Strehl ratios for PALM-3000 and PULSE for different magnitudes of natural guide star. Five different curves are shown in the plot: PALM-3000 s64 wavefront sensor mode (8.2 cm subapertures); PALM-3000 s8 wavefront sensor mode (63.8 cm subapertures); PALM-3000 s8 mode, but with the effects of fitting error removed; the hard performance limit imposed using traditional Rayleigh LGS (§3.1); and PULSE performance using our combination of a UV Rayleigh LGS wavefront sensor having 12.8 cm subapertures (40 subapertures across the telescope aperture) and our NGS wavefront sensor with 63.8 cm subapertures. Each point assumes an optimized sensor frame rate.

**3.3 Wavefront Sensor Packaging and Wavefront Control Strategy**

The near-field nature of a 10 km Rayleigh LGS used with our 0.45 km focal length telescope results in considerable axial focus difference between the LGS and NGS/science focal plane locations (approximately 1 meter in our F/15 Cassegrain beam). This suggests the LGS and NGS beam footprints on AO optics would differ by up to 2.5 inches in diameter, leading to vignetting without costly replacement of PALM-3000's off-axis parabolas. Low UV transmission from sub-optimal AO coatings and knock-on downstream packaging issues for all our science instruments would be similarly challenging. We choose, therefore, to split off the UV LGS light ahead of the main PULSE optical relay into a separate wavefront sensor arm optimized explicitly for UV transmission. This results in an open-loop wavefront sensing architecture, requiring high dynamic range, linear wavefront sensing of the full atmospheric wavefront error. Such open-loop wavefront sensing has been successfully demonstrated[30] by the ViLLaGEs AO system at Lick Observatory to a wavefront accuracy of 40 nm RMS[31], representing a small term in our overall error budget, as well as within the vision science community[32].

Importantly, our hybrid use of NGS sensing for ~ 50 atmospheric spatial modes, and the natural blindness of our LGS beacon to global tip-tilt errors, means that the linearity requirement on PULSE open-loop control is significantly mitigated even compared to the ViLLaGEs experience. Because the low-order modes are being sensing in closed-loop with NGS, only turbulence errors outside the NGS spatial frequency spectrum need be corrected with good linearity and low hysteresis. The RMS wavefront error contained in turbulence above Zernike mode value N is given by[33],

$$\sigma_{atm} = 0.54\, N^{-\sqrt{3}/4}(D/r_0)^{5/6} \qquad (2)$$

where D is the telescope diameter is $r_0$ is the atmospheric coherence parameter. For PULSE, $\sigma_{atm}$ ~ 224 nm RMS, so that in order to maintain open-loop accuracy to better than 45 nm RMS (our allocation from a more detailed error budget), we require approximately 20% accuracy in our wavefront measurement and control. Using a Shack-Hartmann sensor and sampling each LGS subimage with 4 x 4 pixels, we can achieve sufficiently sensor linearity with considerable margin[33], even after accounting for ~10% deformable mirror hysteresis[34]. Because this is an open-loop LGS wavefront sensor, used with an extended LGS beacon, architectures such as pyramid or curvature sensors offer no practical advantage[35] and represent greater technical risk than Shack-Hartmann sensing, our LGS WFS baseline.

PULSE implementation, with architecture choices based on a flow-down of initial requirements from our simulation results, is shown conceptually in Figure 4, including a new facility atmospheric dispersion corrector, which will correct 0.5 – 2.2 microns simultaneously to the diffraction-limited tolerance of the 5-meter telescope for 45 deg zenith angle. The most direct implementation of the real-time control law, the one implemented in the simulations of (§3.2), is to drive our high-order DM with the UV WFS signal and our low-order DM with the NGS WFS signal independently. During our preliminary design, we will investigate improvements to this control law integrating optimal modal division of the control signal.

## 4. ADAPTIVE OPTICS TECHNOLOGY HERITAGE

### 4.1 PALM-3000: Exoplanet Adaptive Optics System

PALM-3000 is a second generation astronomical adaptive optics facility instrument for the 5.1-meter Hale telescope at Palomar Observatory[1]. PALM-3000 began operations with its record 3,388 active-actuator deformable mirror on June 20, 2011 and has since proven its capability for both high-contrast infrared and diffraction-limited visible-light observations (e.g., figure 4) with root-mean-squared (RMS) residual wavefront error as low as 140 nm. Observers using the Project 1640 instrument[5] have embarked on a 99 night exoplanet survey of A and F type stars[36], making significant advances in stellar speckle suppression techniques[7,8] and calibrated integral field spectrograph data pipelines[37]. The P1640 coronagraph team obtained spectra of all four known planets orbiting the star HR8799 (see figure 5), the first simultaneous spectroscopic observations of multiple exoplanets[9]. In addition, the P1640 team re-characterized a suspected planetary companion to a well-known bright star, reclassifying the object as a brown dwarf[38].

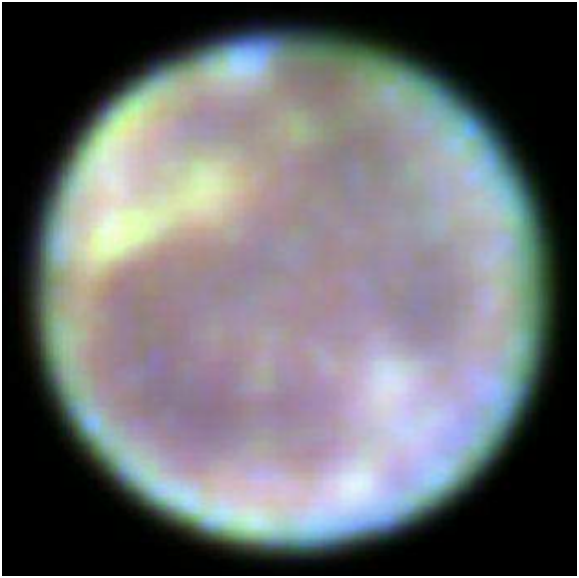 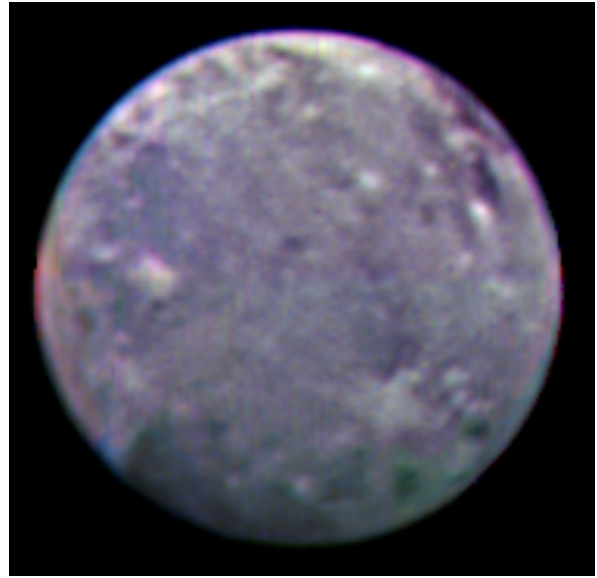

### Hubble Space Telescope             PALM-3000+TMAS

Figure 4. Images of Ganymede with an approximate angular diameter of 1.6 arc seconds. (Left) Hubble Space Telescope false-color image (NASA image). (Right) BRI false color image obtained with PALM-3000 and TMAS[1]. The pixel sampling in this image is 10 milli arc seconds, corresponding to ~ 35 km on the surface of Ganymede at this distance.

PALM-3000 uses two Xinetics, Inc. deformable mirrors in series as a woofer-tweeter pair to correct for measured atmospheric conditions at Palomar Mountain. The low-spatial frequency, large stroke mirror is the original PALM-241[39], which has been in use since 1999. The high-spatial frequency, small stroke mirror is a new 3388 actuator device, the largest format astronomical deformable mirror used on the sky to date[40]. PALM-3000 uses a Shack-Hartmann wavefront sensor with four selectable pupil sampling modes[41] to measure the incoming wavefront. The detector is a

128x128 pixel E2V CCD50 encased in a SciMeasure camera head, originally developed for PALM-241[42]. Pupil sampling modes of 64x, 32x, and 8x subapertures across the entrance pupil allow for performance optimizations on guide stars spanning 18 magnitudes of brightness[43]. Real-time computation is performed by custom software operating on 16 graphics cards distributed over 8 computers, all connected to a servo control and database computer via a high speed network switch[44]. Wavefront reconstruction is optimized on each target by a reconstructor pipeline tool, which automatically gathers wavefront sensor pixel data and weights each controlled actuator based on the flux measured in its matched subaperture. The PALM-3000 operator may optimize the wavefront further by adjusting servo control gains while monitoring telemetry diagnostic displays.

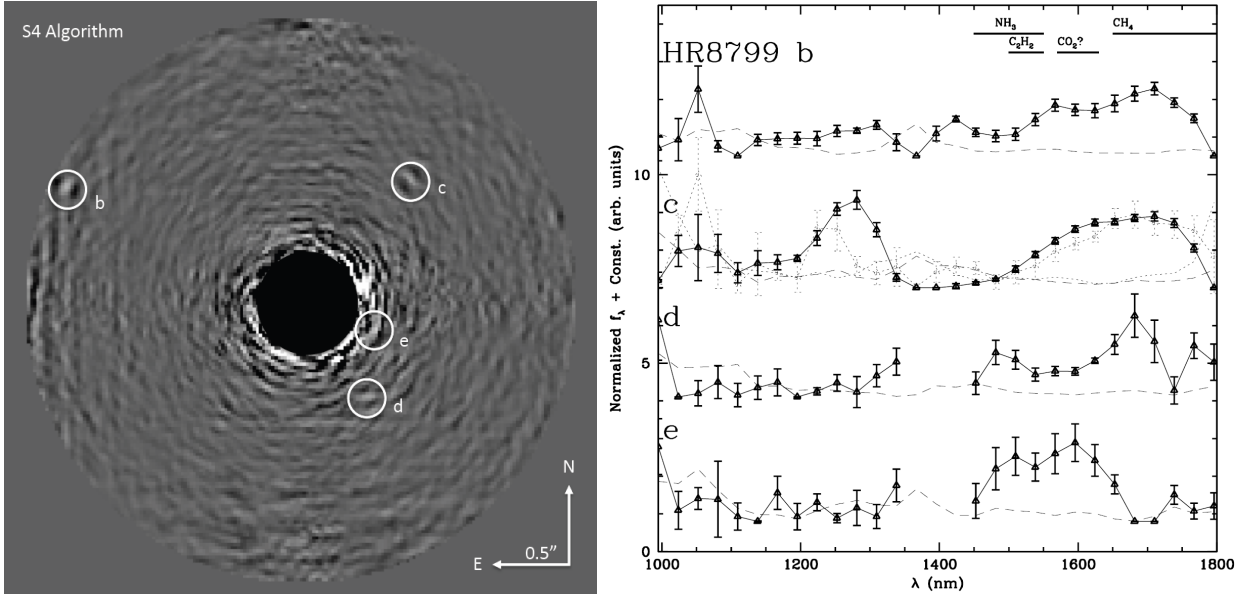

Figure 5. PALM-3000 and Project 1640 collaborators have published 14 journal and conference articles, including the first simultaneous measurement of spectra[9] of all 4 giant planets (b,c,d,e) orbiting HR8799 (left: stacked spectral cube image; right: exoplanet infrared spectra).

### 4.2 Robo-AO: Autonomous Visible-Light Laser Adaptive Optics

To reduce cost and risk, PULSE reuses the UV LGS laser projector and software control of Robo-AO[10,45]. The laser-launch system comprises a pulsed, 12 W, $\lambda$ = 355 nm laser beam, co-aligned with the bore-sight of the principal telescope, and focuses a seeing-limited beam waist with a 15-cm projection aperture to a line-of-sight distance of 10 km. As with other ultraviolet laser systems propagating into navigable airspace[46,47], control measures to avoid illuminating aircraft are not required because the laser beam is unable to flash-blind pilots or produce biologically hazardous radiation levels during momentary exposures. We have measured the Robo-AO projected LGS spot size to be consistently less than $\sqrt{2}$ times the visible seeing conditions, as expected in the absence of tip/tilt information from the LGS. We have measured photoreturn from a 375 m range gate at a zenith distance of 10 km to be ~165 e-/subaperture/exposure at 1.2 kHz sample rate, matching the theoretical return expected from the Lidar equation[33]. As PULSE must coordinate propagation of the UV laser with US Strategic Command to avoid illuminating critical space assets, as with all US LGS AO systems, we will reuse system of tiled predict windows developed for Robo-AO that manages laser deconfliction, allowing us access to the entire sky above 50 degrees zenith distance[48].

The reliability of our Robo-AO elements is demonstrated by over 12,000 scientific observations to date[23], at rates of well over 200 targets per night, with multiple science programs in progress. These include a survey of every Kepler object of interest[49-51]; the Ultimate Binarity Survey to examine stellar binarity properties across the main sequence and beyond[48]; the multiplicity of solar type stars[52]; investigating exotic binary systems[53,54]; probing the properties of unresolved binary star systems[55], and high-precision astrometric observations of the center of M15 to search for a possible intermediate mass black hole.

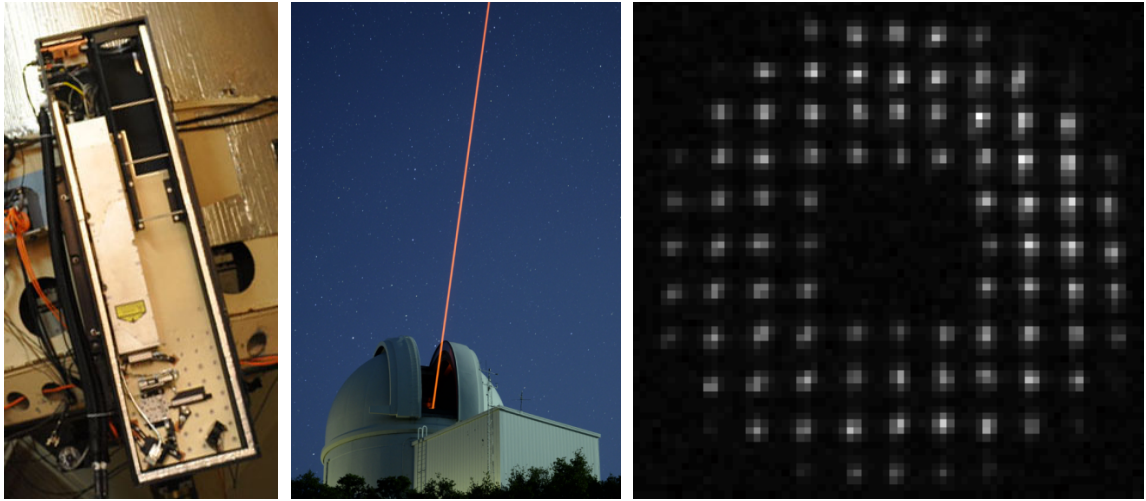

Figure 6. (left) Robo-AO laser projector with cover removed; (center) Robo-AO laser propagating from the 1.5-meter dome (image taken with a UV-sensitive camera); (right) resulting range-gated AO Shack-Hartmann wavefront sensor signal.

## 5. PULSE SCIENCE

### 5.1 Exoplanets in Late-type Stellar Systems

Understanding the architecture of planetary systems across orbital separation and stellar mass is a leading goal of exoplanet science. The statistical properties of planetary architectures provide important clues about formation and migration pathways, but require large surveys using specialized instruments to probe the entire range of planetary orbits (0.1 – 100 AU).

At small separations (<10 AU), radial velocity, transit, and microlensing surveys are showing that planets with masses of 0.1 – 13 $M_{Jup}$ are common products of protoplanetary disk evolution[56]. Beyond 10 AU, high-contrast imaging is the only way to study the exoplanet architectures. At these separations core accretion cannot create gas giant planets (1 – 13 $M_{Jup}$) because orbital periods become prohibitively long. Simulations indicate that the direct collapse of massive protoplanetary disks ("disk instability"), on the other hand, may be a viable formation route in this region[57]. Direct imaging therefore also offers an opportunity to study planets formed entirely differently than those at small separations. The recent discovery via direct imaging of gas giant planets orbiting the high-mass (1.2 – 2.0 $M_{Sun}$) stars HR8799, Fomalhaut, β Pictoris, and GJ504 confirms that giant planets exist at large separations and are within reach of the latest telescope capabilities.

Despite successes, direct imaging programs have mostly neglected low-mass stars. Compared to young (10 – 100 Myr) intermediate- and high-mass stars in the solar neighborhood, few examples of nearby young M dwarfs were known, so past imaging surveys have focused on AFGK-type stars (e.g., The Gemini-NICI Planet-Finding Campaign[58]; The International Deep Planet Search[59]; Strategic Exploration of Exoplanets and Disks with Subaru[60] (SEEDS)). As a result, the statistical properties of wide-separation planets continue to improve for early-type stars but remain inadequate for late-type stars. This selection effect means that giant planets have preferentially been found around high-mass stars, and it is unclear if the intrinsic frequency of wide-separation giant planets varies with stellar host mass. In particular, analysis of short period microlensing events suggests that many M stars may host Jupiter mass planets[61] on wide orbits (3-30 AU) which, for nearby young systems, would be accessible to study with PULSE.

Low-mass stars therefore represent the critical mass range for understanding the influence of stellar host mass on wide-separation planet formation and, because they make up ~75% of stars in our galaxy (Figure 7), are critical to determining the frequency of giant planets across all orbital separations. PULSE will extend the available targets for exoplanet imaging searches at Palomar to include very young (~ 10 Myr) M dwarfs in nearby star-forming regions and young (~100 Myr) M dwarfs within nearby moving groups[62-64] as far as 200 pc. PULSE project scientist Brendan Bowler, has identified a sample of ~1000 nearby young M dwarfs from their X-ray and UV activity levels. They are following them up for spectral typing and radial velocity measurements at Kitt Peak and SOAR. We have also begun vetting the sample of binaries with Robo-AO on the Palomar 60-inch since close stellar binaries within ~ 40 AU appear to be dynamically disruptive to protoplanetary disks, and hence giant planet formation. Together with new group members in the literature, we expect to have a sample of ~500 new, nearby (< 80 pc) young (< 120 Myr) M dwarfs which will form the basis for a large new exoplanet imaging survey to be undertaken with PULSE.

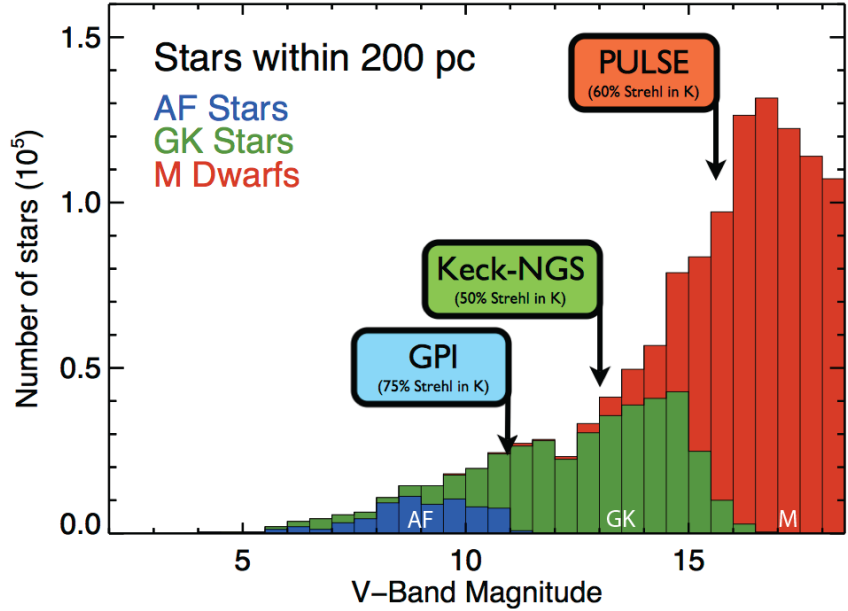

Figure 7. The new parameter space for direct exoplanet study opened with PULSE includes both M dwarfs (mostly unreachable today) and the very youngest early-type stars in associations located up to 200 pc. Star counts are from the TRILEGAL galactic model.

### 5.2 Exoplanets in Young, Early-type Stellar Systems

When planets form, gravitational potential energy is released and turned into tremendous heat. Without an internal nuclear energy source to maintain their temperature, planets cool and dim with time. For massive planets (> $5M_{Jup}$), however, self-luminosity dominates stellar insolation for > $10^8$ years, so that bright giant planets can be found at relatively large orbital separations from young host stars. This advantage was confirmed by the initial direct imaging of the bright young planets at separations > 10AU (β Pic, HR 8799, Fomalhaut). Early-type stars, which are difficult to study with radial velocity or transit photometry due to their activity, currently seem more likely to host giant planets in wide orbits than are FGK-type stars as higher-mass stars may retain large, massive disks that lead to wide-orbit, heavy planets[65].

Using the vector vortex coronagraph within PULSE's PHARO NIR imager[66], thousands of the youngest A and F stars within ~ 200 pc can be observed. At 120 pc in Taurus-Aurigae the < 0.2" inner working angle limit of the vortex allows us to probe < 12 AU from the host.

Using detailed evolutionary models[67], we compute that an age 10 Myr planet of 5 $M_{Jup}$ will have an apparent magnitude of $m_K$ = 15.9, if located at 40 pc. This is well within the demonstrated 5σ sensitivity limits of $m_K$ ~ 16 $hr^{-1/2}$ measured[68] with PHARO at Strehl performance levels half that predicted for PULSE. Comparing these sensitivity curves to the anticipated mass and semi-major axis distribution of exoplanets around solar type stars[69] or slightly more massive stars[62], we expect most young early-type star exoplanet discoveries with PULSE between 40 – 140 pc in orbits 4 – 60 AU region with mass of 3 $M_{Jup}$ or higher. Of specific interest is the case of very young (< 10 Myr) exoplanets, observable before dissipation of the protoplanetary disk, as have been found recently by Spitzer and WISE.

How do these observations provide deeper meaning? The direct detection very young exoplanets reveals their formation process: core accretion, gravitational instability, or a combination of both. Discovery of many giant planets at separation > 10 AU and age < 10 Myr would favor rapid formation through disk accretion. Conversely, it will be hard to reconcile a dearth of planets around stars < 3 Myr with gravitational instability, which predicts planet formation within < 1 Myr. A statistical survey of young exoplanets, only available with high-contrast LGS AO, is critical to understanding planet formation.

## 5.3 Galactic Dark Matter and Gravity in Globular Clusters

PULSE will enable unprecedented ground-based astrometry of globular clusters (GCs) outside the solar circle, based on its combination of faint guide star limit and exquisite correction (and corresponding point spread function stability). High spatial resolution overcomes crowding limits within GC cores and enables the deepest searches for tracers of intermediate mass black holes (IMBHs), when combined with the precision radial velocity information into three-dimensional orbits.

PULSE will also allow us to investigate theoretical post-Newtonian gravity predictions independent of contaminating gravity perturbations in GC outskirts, isolating the truly weak acceleration regime, $a \ll a_0 = 1.2 \times 10^{-10}$ m s$^{-2}$. Alternative theories of gravity[70] can be explored with precise orbits within GCs. One can show that MOND theories lead to different predictions for the dispersion of velocities far from the GC core, 5-10 times the velocity dispersion of Newtonian gravity, corresponding to 10 – 50 μas/yr differential motions[71]. PULSE's visible instruments will be able to overcome crowding limits to determine the long-term dispersion velocities at different core radii, based on HST membership determinations, which can be combined statistically to achieve final precisions between 0.1 and 0.7 km/s depending on GC distance.

Previously published Palomar AO results demonstrated[72] 70 microarcsecond astrometry after 15 minutes of integration without sign of a systematic error floor. Due to improved Strehl and stability, we expect PULSE to achieve < 40 microarcseconds hr$^{-1/2}$ providing significant new proper motion constraints within GCs to complement radial velocity orbits. These measurements can settle the debates regarding IMBHs and the use of GCs as unique laboratories for modified gravity[73].

## 5.4 Galaxy Mergers and Evolution at the Visible Diffraction Limit

The PULSE combination of faint guide star and exquisite AO correction can be singularly exploited by the SWIFT visible-light integral field spectrograph (IFS) for the study of galactic mergers and evolution. An example is Arp 147, a system comprised of a collisionally created ring galaxy and an early-type galaxy. Kinematics studies derived from seeing-limited SWIFT data[74] found the edge-to-edge expansion velocity of the ring is 225 ± 8 km/s, implying an upper limit on the time-scale for the collision of 50 Myr, and demonstrating SWIFT kinematic precision. The ring appears to have been a typical disc galaxy prior to the encounter and shows electron densities consistent with typical values for star-forming HII regions.

With PULSE, such galaxy mergers will be viewable by SWIFT at 8 – 40 times finer spatial resolution than the seeing limit, with excellent red-visible ensquared energy (> 30%) within each of thousands of spatial elements as fine as 16 milliarcsecond. Specifically, we will be able to map the internal kinematics of galaxies to understand their internal evolutionary histories, metallicity and dark matter distributions, and presence of dwarf spheroidal companions. Such a capability is not available in any space telescope, due to lack of instrumentation, nor any other ground-based telescope, due to lack of AO capability at visible red wavelengths using faint guide stars.

Another exciting project ripe for a PULSE + SWIFT is the kinematics of galaxies at redshifts z ~ 1 ([OII] emission lines at 3727 Å and [OIII] lines at 5007 Å allow kinematic studies to z = 1.7 with SWIFT, whose thick fully-depleted CCD detectors have high QE up to 1000 nm). Recent findings[75] show that the kinematics of galaxies change dramatically over the redshift range 1 to 1.5, with higher z objects predominantly disordered, while lower z objects appear to have more ordered kinematics resembling disk galaxies. Is this indeed an early indication of disk formation as the gas dissipates and settles into planar structures?

This is a plausible hypothesis, but one that cannot be tested by 8-10 meter class NIR multi-object spectrographs; though incredibly sensitive, their spatial resolution is limited by atmospheric seeing. Given that the sizes of these objects are typically less than 1", and the velocity field needs to be sampled at size scales of a few kpc, we need AO integral field spectroscopy to correctly measure the kinematic morphology of these objects. While HST observations already show structures on ~ 0.1" spatial scales, these images need to be complemented from the ground with spectroscopy at comparable spatial resolutions. Low sky background in the red visible, relative to the NIR, makes SWIFT the ideal instrument for these studies.

## ACKNOWLEDGMENTS

PALM-3000 was supported by the financial support of National Science Foundation through awards AST-0619922 and AST-1007046, Jet Propulsion Laboratory, Caltech Optical Observatories, and the generous philanthropy of Ron and Glo

Helin. Development of the PALM-3000 high-order deformable mirror was funded by NASA SBIR award #NNG05CA21C. The Robo-AO system is supported by collaborating partner institutions, the California Institute of Technology and the Inter-University Centre for Astronomy and Astrophysics, by the National Science Foundation under Grant Nos. AST-0906060, AST-0960343, and AST-1207891, by a grant from the Mt. Cuba Astronomical Foundation and by a gift from Samuel Oschin. C.B. acknowledges support from the Alfred P. Sloan Foundation.

The successful deployment of PALM-3000 and Robo-AO could not have been possible without the notable talent and dedication of the entire Palomar Observatory staff. We gratefully acknowledge the specific contributions from Dan McKenna, John Henning, Steve Kunsman, Kajsa Peffer, Jean Mueller, Kevin Rykowski, Carolyn Heffner, Greg van Idsinga, Mike Doyle, Bruce Baker, Richard Walters, Karl Dunscombe, and Drew Roderick.